\pdfoutput=1
\documentclass[12pt,a4paper]{article}

\usepackage{ifthen} 
\newboolean{pdflatex}
\setboolean{pdflatex}{true} 

\newboolean{articletitles}
\setboolean{articletitles}{true} 

\newboolean{uprightparticles}
\setboolean{uprightparticles}{false} 

\newboolean{inbibliography}
\setboolean{inbibliography}{false} 

\usepackage{microtype}
\usepackage{lineno}  
\usepackage{xspace} 

\usepackage{graphicx}  
\usepackage{color}
\usepackage{colortbl}
\graphicspath{{./figs/}} 

\usepackage{amsmath} 
\usepackage{amssymb}
\usepackage{amsfonts}
\usepackage{upgreek} 

\newcommand*\patchAmsMathEnvironmentForLineno[1]{%
\expandafter\let\csname old#1\expandafter\endcsname\csname #1\endcsname
\expandafter\let\csname oldend#1\expandafter\endcsname\csname
end#1\endcsname
 \renewenvironment{#1}%
   {\linenomath\csname old#1\endcsname}%
   {\csname oldend#1\endcsname\endlinenomath}%
}
\newcommand*\patchBothAmsMathEnvironmentsForLineno[1]{%
  \patchAmsMathEnvironmentForLineno{#1}%
  \patchAmsMathEnvironmentForLineno{#1*}%
}
\AtBeginDocument{%
\patchBothAmsMathEnvironmentsForLineno{equation}%
\patchBothAmsMathEnvironmentsForLineno{align}%
\patchBothAmsMathEnvironmentsForLineno{flalign}%
\patchBothAmsMathEnvironmentsForLineno{alignat}%
\patchBothAmsMathEnvironmentsForLineno{gather}%
\patchBothAmsMathEnvironmentsForLineno{multline}%
}

\usepackage{hyperref}    
\usepackage[all]{hypcap} 

\def\lhcb {\mbox{LHCb}\xspace}
\def\ux85 {\mbox{UX85}\xspace}

\ifthenelse{\boolean{uprightparticles}}%
{

 \def\Peta        {\ensuremath{\upeta}\xspace}

 \def\Pmu         {\ensuremath{\upmu}\xspace}

 \def\Ppi         {\ensuremath{\uppi}\xspace}

 \def\Ppsi        {\ensuremath{\uppsi}\xspace}

 \def\PDelta      {\ensuremath{\Delta}\xspace}                 
 \def\PXi      {\ensuremath{\Xi}\xspace}                 
 \def\PLambda      {\ensuremath{\Lambda}\xspace}                 
 \def\PSigma      {\ensuremath{\Sigma}\xspace}                 
 \def\POmega      {\ensuremath{\Omega}\xspace}                 
 \def\PUpsilon      {\ensuremath{\Upsilon}\xspace}

 \def\PB      {\ensuremath{\mathrm{B}}\xspace}                 
                  
 \def\PD      {\ensuremath{\mathrm{D}}\xspace}

 \def\PJ      {\ensuremath{\mathrm{J}}\xspace}                 
 \def\PK      {\ensuremath{\mathrm{K}}\xspace}

 \def\Pb      {\ensuremath{\mathrm{b}}\xspace}                 
 \def\Pc      {\ensuremath{\mathrm{c}}\xspace}                 
 \def\Pd      {\ensuremath{\mathrm{d}}\xspace}

 \def\Pi      {\ensuremath{\mathrm{i}}\xspace}

 \def\Pp      {\ensuremath{\mathrm{p}}\xspace}                 
 \def\Pq      {\ensuremath{\mathrm{q}}\xspace}                 
                  
 \def\Ps      {\ensuremath{\mathrm{s}}\xspace}                 
                  
 \def\Pu      {\ensuremath{\mathrm{u}}\xspace}

}
{

 \def\Peta        {\ensuremath{\eta}\xspace}

 \def\Pmu         {\ensuremath{\mu}\xspace}

 \def\Ppi         {\ensuremath{\pi}\xspace}

 \def\Ppsi        {\ensuremath{\psi}\xspace}                 
                  
 \mathchardef\PDelta="7101
 \mathchardef\PXi="7104
 \mathchardef\PLambda="7103
 \mathchardef\PSigma="7106
 \mathchardef\POmega="710A
 \mathchardef\PUpsilon="7107
                  
 \def\PB      {\ensuremath{B}\xspace}                 
                  
 \def\PD      {\ensuremath{D}\xspace}

 \def\PJ      {\ensuremath{J}\xspace}                 
 \def\PK      {\ensuremath{K}\xspace}

 \def\Pb      {\ensuremath{b}\xspace}                 
 \def\Pc      {\ensuremath{c}\xspace}                 
 \def\Pd      {\ensuremath{d}\xspace}

 \def\Pi      {\ensuremath{i}\xspace}

 \def\Pp      {\ensuremath{p}\xspace}                 
 \def\Pq      {\ensuremath{q}\xspace}                 
                  
 \def\Ps      {\ensuremath{s}\xspace}                 
                  
 \def\Pu      {\ensuremath{u}\xspace}

}

\def\mumu       {\ensuremath{\Pmu^+\Pmu^-}\xspace}

\def\quark     {\ensuremath{\Pq}\xspace}

\def\uquark    {\ensuremath{\Pu}\xspace}

\def\dquark    {\ensuremath{\Pd}\xspace}

\def\squark    {\ensuremath{\Ps}\xspace}

\def\cquark    {\ensuremath{\Pc}\xspace}

\def\bquark    {\ensuremath{\Pb}\xspace}

\def\pion  {\ensuremath{\Ppi}\xspace}

\def\pip   {\ensuremath{\pion^+}\xspace}
\def\pim   {\ensuremath{\pion^-}\xspace}

\def\kaon  {\ensuremath{\PK}\xspace}
  \def\Kbar  {\kern 0.2em\overline{\kern -0.2em \PK}{}\xspace}

\def\Kz    {\ensuremath{\kaon^0}\xspace}
\def\Kzb   {\ensuremath{\Kbar^0}\xspace}
\def\KzKzb {\ensuremath{\Kz \kern -0.16em \Kzb}\xspace}
\def\Kp    {\ensuremath{\kaon^+}\xspace}
\def\Km    {\ensuremath{\kaon^-}\xspace}

\def\KpKm  {\ensuremath{\Kp \kern -0.16em \Km}\xspace}

\newcommand{\etapr}{\ensuremath{\Peta^{\prime}}\xspace}

  \def\Dbar    {\kern 0.2em\overline{\kern -0.2em \PD}{}\xspace}
\def\D       {\ensuremath{\PD}\xspace}
\def\Db      {\ensuremath{\Dbar}\xspace}
\def\Dz      {\ensuremath{\D^0}\xspace}
\def\Dzb     {\ensuremath{\Dbar^0}\xspace}
\def\DzDzb   {\ensuremath{\Dz {\kern -0.16em \Dzb}}\xspace}
\def\Dp      {\ensuremath{\D^+}\xspace}
\def\Dm      {\ensuremath{\D^-}\xspace}

\def\DpDm    {\ensuremath{\Dp {\kern -0.16em \Dm}}\xspace}

\def\B       {\ensuremath{\PB}\xspace}
\def\Bbar    {\ensuremath{\kern 0.18em\overline{\kern -0.18em \PB}{}}\xspace}

\def\Bz      {\ensuremath{\B^0}\xspace}

\def\Bu      {\ensuremath{\B^+}\xspace}

\def\Bp      {\ensuremath{\Bu}\xspace}

\def\Bd      {\ensuremath{\B^0}\xspace}
\def\Bs      {\ensuremath{\B^0_\squark}\xspace}

\def\Bds     {\ensuremath{\B^0_{(\squark)}}\xspace}

\def\jpsi     {\ensuremath{{\PJ\mskip -3mu/\mskip -2mu\Ppsi\mskip 2mu}}\xspace}

\def\Y#1S{\ensuremath{\PUpsilon{(#1S)}}\xspace}

\def\proton      {\ensuremath{\Pp}\xspace}
\def\antiproton  {\ensuremath{\overline \proton}\xspace}
\def\pbar        {\antiproton}
\def\ppbar       {\ensuremath{\proton\pbar}\xspace}

\def\L {\ensuremath{\PLambda}\xspace}
\def\Lbar {\ensuremath{\kern 0.1em\overline{\kern -0.1em\PLambda}}\xspace}

\newcommand{\decay}[2]{\ensuremath{#1\!\to #2}\xspace}       

\def\to                 {\ensuremath{\rightarrow}\xspace}

\def\BToJpsipp   {\decay{\Bds}{\jpsi\proton\antiproton}}
\def\BdToJpsipp   {\decay{\Bd}{\jpsi\proton\antiproton}}
\def\BsToJpsipp   {\decay{\Bs}{\jpsi\proton\antiproton}}
\def\BuToJpsipppi   {\decay{\Bp}{\jpsi\proton\antiproton\pip}}
\def\BToJpsipipi   {\decay{\Bds}{\jpsi\pip\pim}}
\def\BsToJpsipipi   {\decay{\Bs}{\jpsi\pip\pim}}
\def\BdToJpsipipi   {\decay{\Bd}{\jpsi\pip\pim}}

\def\BdToJpsiKpi  {\decay{\Bd}{\jpsi \Kp \pim}}

\def\AT#1     {\ensuremath{A_{\mathrm{T}}^{#1}}\xspace}

\def\C#1      {\ensuremath{\mathcal{C}_{#1}}\xspace}                     
\def\Cp#1     {\ensuremath{\mathcal{C}_{#1}^{'}}\xspace}                 
\def\Ceff#1   {\ensuremath{\mathcal{C}_{#1}^{\mathrm{(eff)}}}\xspace}    
\def\Cpeff#1  {\ensuremath{\mathcal{C}_{#1}^{'\mathrm{(eff)}}}\xspace}   
\def\Ope#1    {\ensuremath{\mathcal{O}_{#1}}\xspace}                     
\def\Opep#1   {\ensuremath{\mathcal{O}_{#1}^{'}}\xspace}

\newcommand{\tev}{\ifthenelse{\boolean{inbibliography}}{\ensuremath{~T\kern -0.05em eV}\xspace}{\ensuremath{\mathrm{\,Te\kern -0.1em V}}\xspace}}
\newcommand{\gev}{\ensuremath{\mathrm{\,Ge\kern -0.1em V}}\xspace}
\newcommand{\mev}{\ensuremath{\mathrm{\,Me\kern -0.1em V}}\xspace}
\newcommand{\kev}{\ensuremath{\mathrm{\,ke\kern -0.1em V}}\xspace}
\newcommand{\ev}{\ensuremath{\mathrm{\,e\kern -0.1em V}}\xspace}
\newcommand{\gevc}{\ensuremath{{\mathrm{\,Ge\kern -0.1em V\!/}c}}\xspace}
\newcommand{\mevc}{\ensuremath{{\mathrm{\,Me\kern -0.1em V\!/}c}}\xspace}
\newcommand{\gevcc}{\ensuremath{{\mathrm{\,Ge\kern -0.1em V\!/}c^2}}\xspace}
\newcommand{\gevgevcccc}{\ensuremath{{\mathrm{\,Ge\kern -0.1em V^2\!/}c^4}}\xspace}
\newcommand{\mevcc}{\ensuremath{{\mathrm{\,Me\kern -0.1em V\!/}c^2}}\xspace}

\def\mm   {\ensuremath{\rm \,mm}\xspace}

\def\mum  {\ensuremath{\,\upmu\rm m}\xspace}

\def\invfb   {\ensuremath{\mbox{\,fb}^{-1}}\xspace}

\newcommand{\stat}{\ensuremath{\mathrm{\,[stat]}}\xspace}
\newcommand{\syst}{\ensuremath{\mathrm{\,[syst]}}\xspace}

\newcommand{\chisq}{\ensuremath{\chi^2}\xspace}

\newcommand{\chisqvtx}{\ensuremath{\chi^2_{\rm vtx}}\xspace}
\newcommand{\chisqip}{\ensuremath{\chi^2_{\rm IP}}\xspace}

\def\gsim{{~\raise.15em\hbox{$>$}\kern-.85em
          \lower.35em\hbox{$\sim$}~}\xspace}
\def\lsim{{~\raise.15em\hbox{$<$}\kern-.85em
          \lower.35em\hbox{$\sim$}~}\xspace}

\def\sPlot{\mbox{\em sPlot}\xspace}

\def\pt         {\mbox{$p_{\rm T}$}\xspace}

\def\evtgen     {\mbox{\textsc{EvtGen}}\xspace}
\def\pythia     {\mbox{\textsc{Pythia}}\xspace}

\def\geant      {\mbox{\textsc{Geant4}}\xspace}

\def\photos     {\mbox{\textsc{Photos}}\xspace}

\def\pythia     {\mbox{\textsc{Pythia}}\xspace}

\def\tell1  {TELL1\xspace}
\def\ukl1   {UKL1\xspace}

\usepackage{cite} 
\usepackage{mciteplus}

\begin{document}

\renewcommand{\thefootnote}{\fnsymbol{footnote}}
\setcounter{footnote}{1}

\begin{titlepage}
\pagenumbering{roman}

\vspace*{-1.5cm}
\centerline{\large EUROPEAN ORGANIZATION FOR NUCLEAR RESEARCH (CERN)}
\vspace*{1.5cm}
\hspace*{-0.5cm}
\begin{tabular*}{\linewidth}{lc@{\extracolsep{\fill}}r}
\ifthenelse{\boolean{pdflatex}}
{\vspace*{-2.7cm}\mbox{\!\!\!\includegraphics[width=.14\textwidth]{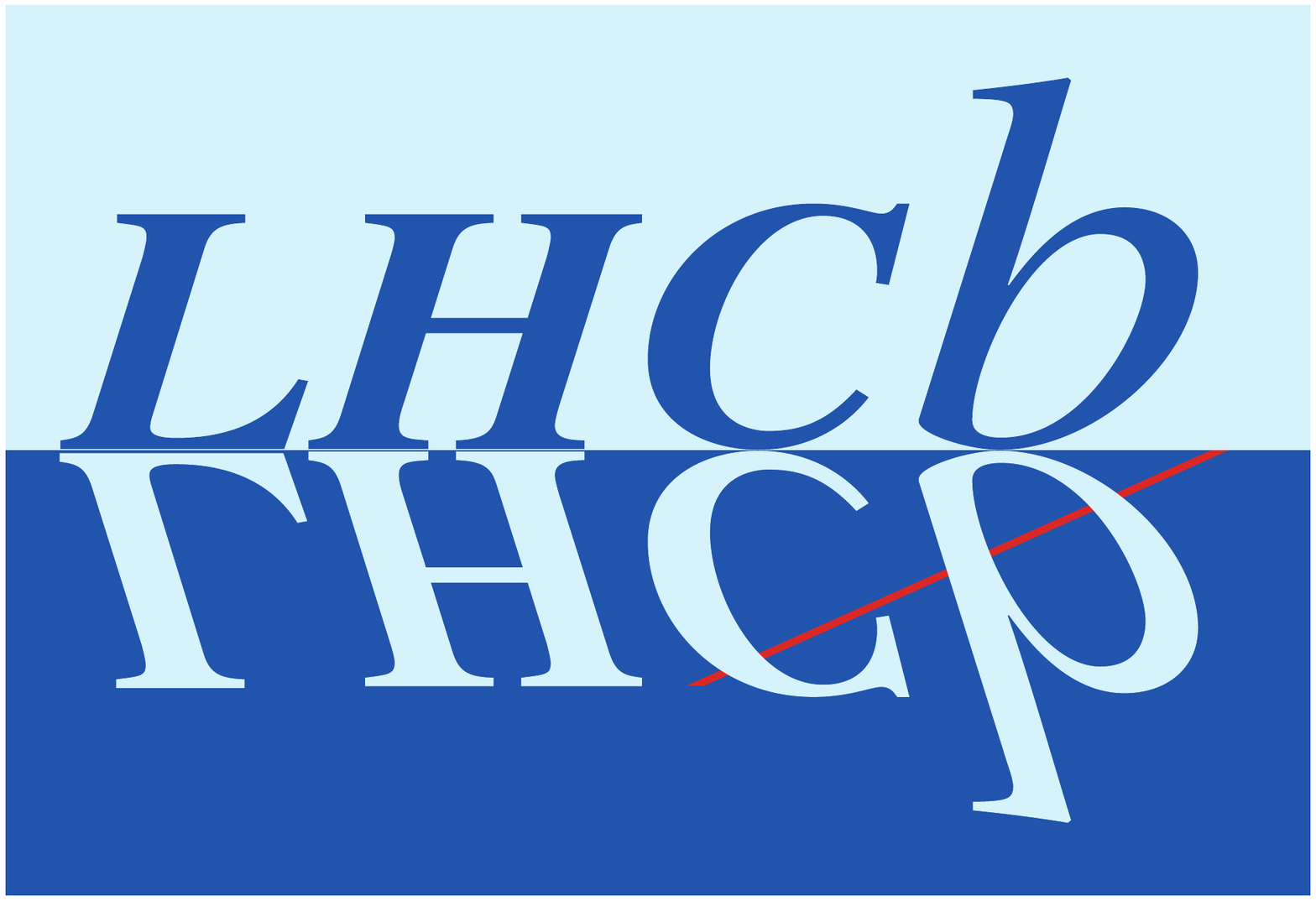}} & &}
{\vspace*{-1.2cm}\mbox{\!\!\!\includegraphics[width=.12\textwidth]{figs/lhcb-logo.eps}} & &}
\\
 & & CERN-PH-EP-2013-099 \\  
 & & LHCb-PAPER-2013-029 \\ 
 & & July 25, 2013 \\ 
 & & \\
\end{tabular*}

\vspace*{2.0cm}

{\bf\boldmath\huge
\begin{center}
  Searches for $\Bds \to \jpsi \ppbar$ and $\Bp \to \jpsi \ppbar \pip$ decays
\end{center}
}

\vspace*{1.0cm}

\begin{center}
The LHCb collaboration\footnote{Authors are listed on the following pages.}
\end{center}

\vspace{\fill}

\begin{abstract}
  \noindent
  The results of searches for $\Bds \to \jpsi \ppbar$ and $\Bp \to \jpsi \ppbar \pip$ decays are reported.
  The analysis is based on a data sample, corresponding to an integrated luminosity of $1.0 \invfb$ of $pp$ collisions,  
  collected with the LHCb detector.
  An excess with 2.8\,$\sigma$ significance is seen for the decay \BsToJpsipp and an upper limit on the branching
  fraction is set at the 90\,\% confidence level: ${\cal B}(\Bs \to \jpsi \ppbar) < 4.8\times 10^{-6}$, 
  which is the first such limit.
  No significant signals are seen for \BdToJpsipp and \BuToJpsipppi decays, for which the corresponding limits are set:
  ${\cal B}(\Bd \to \jpsi \ppbar) < 5.2\times 10^{-7}$, which significantly improves the existing limit; 
  and ${\cal B}(\Bp \to \jpsi \ppbar \pip) < 5.0\times 10^{-7}$,
  which is the first limit on this branching fraction.
\end{abstract}

\vspace*{1.0cm}

\begin{center}
  Submitted to JHEP
\end{center}

\vspace{\fill}

{\footnotesize 
\centerline{\copyright~CERN on behalf of the \lhcb collaboration, license \href{http://creativecommons.org/licenses/by/3.0/}{CC-BY-3.0}.}}
\vspace*{2mm}

\end{titlepage}

\newpage
\setcounter{page}{2}
\mbox{~}
\newpage

%%%%%%%%%%%%%%%%%%%%%%%%%%%%%%%%%%%%%%%%%%
\centerline{\large\bf LHCb collaboration}
\begin{flushleft}
\small
R.~Aaij$^{40}$, 
B.~Adeva$^{36}$, 
M.~Adinolfi$^{45}$, 
C.~Adrover$^{6}$, 
A.~Affolder$^{51}$, 
Z.~Ajaltouni$^{5}$, 
J.~Albrecht$^{9}$, 
F.~Alessio$^{37}$, 
M.~Alexander$^{50}$, 
S.~Ali$^{40}$
% ...

{\footnotesize \it
% ...
$ ^{5}$Clermont Universit\'{e}, Universit\'{e} Blaise Pascal, CNRS/IN2P3, LPC, Clermont-Ferrand, France\\
$ ^{6}$CPPM, Aix-Marseille Universit\'{e}, CNRS/IN2P3, Marseille, France\\
% ...
$ ^{9}$Fakult\"{a}t Physik, Technische Universit\"{a}t Dortmund, Dortmund, Germany\\
% ...
$ ^{36}$Universidad de Santiago de Compostela, Santiago de Compostela, Spain\\
$ ^{37}$European Organization for Nuclear Research (CERN), Geneva, Switzerland\\
% ...
$ ^{40}$Nikhef National Institute for Subatomic Physics, Amsterdam, The Netherlands\\
% ...
$ ^{45}$H.H. Wills Physics Laboratory, University of Bristol, Bristol, United Kingdom\\
% ...
$ ^{50}$School of Physics and Astronomy, University of Glasgow, Glasgow, United Kingdom\\
$ ^{51}$Oliver Lodge Laboratory, University of Liverpool, Liverpool, United Kingdom\\
% ...
}
\end{flushleft}
%%%%%%%%%%%%%%%%%%%%%%%%%%%%%%%%%%%%%%%%%%

\cleardoublepage

\renewcommand{\thefootnote}{\arabic{footnote}}
\setcounter{footnote}{0}

\pagestyle{plain}
\setcounter{page}{1}
\pagenumbering{arabic}

\section{Introduction}

The production of baryon-antibaryon pairs in $B$ meson decays is of significant experimental and theoretical interest.
For example, in the case of $\proton \antiproton$ pair production, the observed decays $\Bd \to \Db^{(*)0}\proton \antiproton$~\cite{Abe:2002tw,delAmoSanchez:2011gi}, $\Bp \to K^{(*)+} \proton \antiproton$~\cite{Aubert:2005gw,Aubert:2007qea,Wei:2007fg,Chen:2008jy,LHCb-PAPER-2012-047}, $\Bz \to K^{(*)0} \proton \antiproton$~\cite{Aubert:2007qea,Chen:2008jy} and $\Bp \to \pip \proton \antiproton$~\cite{Aubert:2007qea,Wei:2007fg} all have an enhancement near the $\proton \antiproton$ threshold.\footnote{Throughout this paper, the inclusion of charge-conjugate processes is implied.
}
Possible explanations for this behaviour include the existence of an intermediate state in the  $\proton \antiproton$ system~\cite{Chua:2002wp} and short-range correlations between $\proton$ and $\antiproton$ in their fragmentation~\cite{Chua:2002wn,Kerbikov:2004gs,Cheng:2001tr}.
Moreover, for each of these decays, the branching fraction is approximately $10\,\%$ that of the corresponding decay with $\proton \antiproton$ replaced by $\pip\pim$~\cite{PDG2012}.
In contrast, the decay $\Bd \to \jpsi \proton \antiproton$ has not yet been observed; the most restrictive upper limit being ${\cal B}(\Bd \to \jpsi \proton \antiproton) < 8.3 \times 10^{-7}$ at 90\,\% confidence level~\cite{Xie:2005tf}, approximately fifty times lower than the branching fraction for $\Bd \to \jpsi \pip\pim$ decays~\cite{LHCb-PAPER-2012-045}.
This result is in tension with the theoretical prediction of ${\cal B}(\Bd \to \jpsi \proton \antiproton) = (1.2 \pm 0.2) \times 10^{-6}$~\cite{Chen:2008sw}.
Improved experimental information on the $\Bd \to \jpsi \proton \antiproton$ decay would help to understand the process of dibaryon production.

In this paper, the results of a search for \BdToJpsipp and \BsToJpsipp decays are presented.
No prediction or experimental limit exists for the branching fraction ${\cal B}(\Bs \to \jpsi \proton\antiproton)$,
but it is of interest to measure the suppression relative to $\Bs \to \jpsi \pip\pim$~\cite{LHCb-PAPER-2012-005}.
In addition, a search for the decay $\Bp \to \jpsi \proton \antiproton \pip$ is performed,
for which no published measurement exists.
All branching fractions are measured relative to that of the decay
\BsToJpsipipi, which is well suited for this purpose due to its similar
topology to the signal decays.
Additionally, the lower background level and its more precisely 
measured branching fraction make it a more suitable normalisation channel than the companion \Bz mode.

\section{Detector and dataset}

The \lhcb detector~\cite{Alves:2008zz} is a single-arm forward
spectrometer covering the \mbox{pseudorapidity} range $2<\eta <5$,
designed for the study of particles containing \bquark or \cquark
quarks. The detector includes a high precision tracking system
consisting of a silicon-strip vertex detector surrounding the $pp$
interaction region, a large-area silicon-strip detector located
upstream of a dipole magnet with a bending power of about
$4{\rm\,Tm}$, and three stations of silicon-strip detectors and straw
drift tubes placed downstream.
The combined tracking system provides momentum measurement with
relative uncertainty that varies from 0.4\% at 5\gevc to 0.6\% at 100\gevc, 
and impact parameter (IP) resolution of 20\mum for
tracks with high transverse momentum (\pt). 
Charged hadrons are identified using two ring-imaging Cherenkov detectors~\cite{LHCb-DP-2012-003}. 
Photon, electron and hadron candidates are identified by a calorimeter system consisting of
scintillating-pad and preshower detectors, an electromagnetic
calorimeter and a hadronic calorimeter. Muons are identified by a
system composed of alternating layers of iron and multiwire
proportional chambers~\cite{LHCb-DP-2012-002}. 
The trigger~\cite{LHCb-DP-2012-004} consists of a
hardware stage, based on information from the calorimeter and muon
systems, followed by a software stage, which applies a full event
reconstruction.

The analysis uses a data sample, corresponding to an integrated luminosity
of $1.0\invfb$ of $pp$ collision data at a centre-of-mass energy of $7\tev$,
collected with the LHCb detector during 2011.
Samples of simulated events are also used to determine the signal
selection efficiency, to model signal event distributions and to
investigate possible background contributions.
In the simulation, $pp$ collisions are generated using
\pythia~6.4~\cite{Sjostrand:2006za} with a specific \lhcb
configuration~\cite{LHCb-PROC-2010-056}.  Decays of hadronic particles
are described by \evtgen~\cite{Lange:2001uf}, in which final state
radiation is generated using \photos~\cite{Golonka:2005pn}. The
interaction of the generated particles with the detector and its
response are implemented using the \geant
toolkit~\cite{Allison:2006ve, *Agostinelli:2002hh} as described in
Ref.~\cite{LHCb-PROC-2011-006}.

\section{Trigger and selection requirements}

The trigger requirements for this analysis exploit the signature of the
$\jpsi\to\mumu$ decay, and hence are the same for the signal and the \BsToJpsipipi control channel.
At the hardware stage either one or two identified muon candidates are required.
In the case of single muon triggers, the transverse momentum of the candidate is required to be larger than $1.5\gevc$.  
For dimuon candidates a requirement on the product of the \pt~of the muon candidates is applied,  \mbox{$\sqrt{\pt_1 \pt_2} > 1.3\gevc$}. 
In the subsequent software trigger, at least one of the final state muons
is required to have both $\pt>1.0\gevc$ and ${\rm IP}>100\mum$.
Finally, the muon tracks are required to form a vertex that is
significantly displaced from the primary vertices (PVs) and to have invariant mass within
$120\mevcc$ of the known \jpsi mass, $m_\jpsi$~\cite{PDG2012}.

The selection uses a multivariate algorithm (hereafter referred to as MVA) to reject background.
A neural network is trained on data using the \BsToJpsipipi control channel as a proxy for the signal decays.
Preselection criteria are applied in order to obtain a clean sample of the
control channel decays.
The muons from the \jpsi decay must be well identified and have $\pt > 500\mevc$.  
They should also form a vertex with $\chisqvtx < 12$ and have invariant mass within the range $-48 < m_{\mumu} - m_{\jpsi} < 43 \mevcc$.
The separation of the \jpsi vertex from all PVs must be greater than $3\mm$.
The pion candidates must be inconsistent with the muon hypothesis, have $\pt > 200\mevc$ and have minimum \chisqip with respect to any of the PVs greater than 9, 
where the \chisqip is defined as the difference in \chisq of a given PV reconstructed with and without the considered track.
In addition, the scalar sum of their transverse momenta must be greater than
$600\mevc$.
The \B candidate formed from the \jpsi and two oppositely charged hadron
candidates should have $\chisqvtx < 20$ and a minimum \chisqip with
respect to any of the PVs less than 30.
In addition, the cosine of the angle between the \B candidate momentum
vector and the line joining the associated PV and the \B decay vertex
(\B pointing angle) should be greater than 0.99994.

The mass distribution of candidate \BToJpsipipi decays remaining after the preselection is then fitted in order to obtain signal and background distributions of the variables that enter the MVA training, using the \sPlot\ technique~\cite{Pivk:2004ty}.
The fit model is described in Sec.~\ref{sec:fit}.
The variables that enter the MVA training are chosen to minimise any
difference in the selection between the signal and control channels.  
Different selection algorithms are trained for the \BToJpsipp mode and for the
\BuToJpsipppi mode, with slightly different sets of variables.
The variables in common between the selections are the minimum \chisqip
of the \B candidate; the cosine of the \B pointing angle; the \chisq of the
\B and \jpsi candidate vertex fits; the \chisq per degree of freedom of the
track fit of the charged hadrons; and the minimum IP of the muon candidates.
For the \BToJpsipp selection the following additional variables are included:
the \pt of the charged hadron and \jpsi candidates;  
the \pt of the \B candidate; and the flight distance and flight 
distance significance squared of the \B candidate from its associated PV. 
For the \BuToJpsipppi selection only the momentum and
\pt of the muon candidates are included as additional variables.

The MVAs are trained using the {\mbox{\textsc{NeuroBayes}}\xspace} package~\cite{Feindt:2006pm}.
Two different figures of merit are considered to find the optimal MVA requirement.
The first is that suggested in Ref.~\cite{Punzi}
\begin{equation}
\label{eq:punzi}
{\cal{Q}}_1 = \frac{\epsilon_{\rm MVA}} {{a/2} + \sqrt{B_{\rm MVA}}}\,,
\end{equation}
where $a=3$ and quantifies the target level of significance, 
$\epsilon_{\rm MVA}$ is the efficiency of the selection of the signal
candidates, which is determined from simulated signal samples, and
$B_{\rm MVA}$ is the expected number of background events in the signal region; 
which is estimated by performing a fit to the invariant mass distribution of the
data sidebands. 
The second figure of merit is an estimate of the 
expected 90\,\% confidence level upper limit on the branching fraction in the
case that no signal is observed
\begin{equation}
\label{eq:ul-fom}
{\cal{Q}}_2 = \frac{1.64\,\sigma_{N_{\rm sig}}}{\epsilon_{\rm MVA}}\,,
\end{equation}
where $\sigma_{N_{\rm sig}}$ is the expected uncertainty on the signal
yield, which is estimated from pseudo-experiments generated with the
background-only hypothesis.
The maximum of the first and the minimum of the second figure of merit are found to occur at very similar values.
For the \BToJpsipp (\BuToJpsipppi) decay, requirements are chosen such that 
approximately 50\,\% (99\,\%) of the signal is retained while reducing the background to 20\,\% (70\,\%) of its level prior to the cut.
The background level for the \BuToJpsipppi decay is very low due to its proximity to threshold, and only a loose MVA requirement is necessary.

The particle identification (PID) selection for the signal modes is optimised in a similar way using Eq.~(\ref{eq:punzi}).  
It is found that, for the signal channels, placing a tight requirement on the  
proton with a higher value for the logarithm of the likelihood ratio of the proton and pion hypotheses~\cite{LHCb-DP-2012-003} 
and a looser requirement on the other proton results in much better performance than applying the same requirement on both protons.
No PID requirements are made on the pion track in the \BuToJpsipppi mode.

The acceptance and selection efficiencies are determined from simulated
signal samples, except for those of the PID requirements, 
which are determined from data control samples to avoid biases due to known discrepancies between data and simulation.
High-purity control samples of $\L \to p\pim$ ($\Dz \to \Km\pip$) decays with no PID selection requirements applied are
used to tabulate efficiencies for protons (pions) as a function of their momentum and \pt.  
The kinematics of the simulated signal events are
then used to determine an average efficiency.
Possible variations of the efficiencies over the multibody phase space
are considered.  The efficiencies are determined in bins of the Dalitz plot, 
$m^2_{\jpsi h^+}$ vs. $m^2_{h^+ h^-}$, where $h = \pi, \proton$; 
the \jpsi decay angle (defined as the angle between the $\mu^{+}$ and the $\proton\antiproton$ system in the \jpsi rest frame); and the angle between the
decay planes of the \jpsi and the $h^+ h^-$ system.
The variation with the Dalitz plot variables is the most significant.
For the \BsToJpsipipi control sample, the distribution of the signal in
the phase space variables is determined using the \sPlot technique
and these distributions are used to find a weighted average efficiency.

A number of possible background modes, such as cross-feed from  
$\Bds \to \jpsi h^{+} h^{\prime -}$ final states (where $h^{(\prime)} = \pi, K$), 
have been studied using simulation.  
None of these are found to give a significant peaking contribution to the \B candidate invariant mass distribution 
once all the selection criteria had been applied.
Therefore, all backgrounds in the fits to the mass distributions of \BToJpsipp and \BuToJpsipppi candidates are considered as being combinatorial in nature.  
For the fits to the \BsToJpsipipi control channel, some particular backgrounds are taken into account, as described in the following section.

After all selection requirements are applied, 854 and 404 candidates are
found in the invariant mass ranges $[5167,5478]\mevcc$ and
$[5129,5429]\mevcc$ for \BToJpsipp and \BuToJpsipppi decays, respectively.
The efficiency ratios, with respect to the \BsToJpsipipi normalisation
channel, including contributions from detector acceptance, trigger and 
selection criteria (but not from PID) are $0.92 \pm 0.16$, $0.85 \pm 0.12$ and $0.17 \pm 0.04$ for 
\BdToJpsipp, \BsToJpsipp and \BuToJpsipppi, respectively.
In addition, the relative PID efficiencies are found to be $0.78 \pm 0.02$, $0.79 \pm 0.02$ 
and $1.00 \pm 0.03$ for \BdToJpsipp, \BsToJpsipp and \BuToJpsipppi, respectively.
The systematic uncertainties arising from these values are discussed in Sec.~\ref{sec:syst}.

\section{Fit model and results}
\label{sec:fit}

Signal and background event yields are estimated by performing unbinned 
extended maximum likelihood fits to the invariant mass distributions of the 
\B candidates.
The signal probability density functions (PDFs) are parametrised as the sum
of two Crystal Ball (CB) functions~\cite{Skwarnicki:1986xj}, where the power law tails are on opposite sides of the peak.
This form is appropriate to describe the asymmetric tails that result from a combination
of the effects of final state radiation and stochastic tracking
imperfections.
The two CB functions are constrained to have the same peak position, equal to the
value fitted in the simulation.
The resolution parameters are allowed to vary within a Gaussian constraint,
with the central value taken from the simulation and scaled by the
ratio of the values found in the control channel data and corresponding simulation.
The proximity to threshold of the signal decays provides a mass resolution of
1--3\mevcc, whereas for the normalisation channel it is 6--9\mevcc.
The tail parameters and the relative normalisation of the two CB functions
are taken from the simulated distributions and fixed for the fits to
data.  

A second-order polynomial function is used to describe the combinatorial
background component in the \BToJpsipp spectrum while an exponential
function is used for the same component in the \BuToJpsipppi and
\BToJpsipipi channels.
The parameters of these functions are allowed to vary in the fits.
There are several specific backgrounds that contribute to the \BToJpsipipi 
invariant mass spectrum~\cite{LHCb-PAPER-2012-045}, which need to be explicitly modelled. 
In particular, the decay ${\decay{\Bd}{\jpsi\Kp\pim}}$, where a kaon is
misidentified as a pion, is modelled by an exponential function.
The yield of this contribution is allowed to vary in order to
enable a better modelling of the background in the low mass region.
Two additional sources of peaking background are considered: partially
reconstructed decays, such as ${\decay{\Bs}{\jpsi\etapr(\rho\gamma)}}$; and decays where an
additional low momentum pion is included from the rest of the event, such as
${\decay{\Bp}{\jpsi\Kp}}$.
Both distributions are fitted with a non-parametric kernel estimation, with shapes fixed from simulation.
The yields of these components are also fixed to values estimated from the
known branching fractions and selection efficiencies evaluated from
simulation.

In order to validate the stability of the fit, a series of 
pseudo-experiments have been generated using the PDFs described above.  
The experiments are conducted for a wide range of generated signal yields.  
No significant bias is observed in any of the simulation
ensembles; any residual bias being accounted for as a source of systematic uncertainty.

\begin{figure}[htbp]
\begin{center}
\includegraphics*[width=0.49\textwidth]{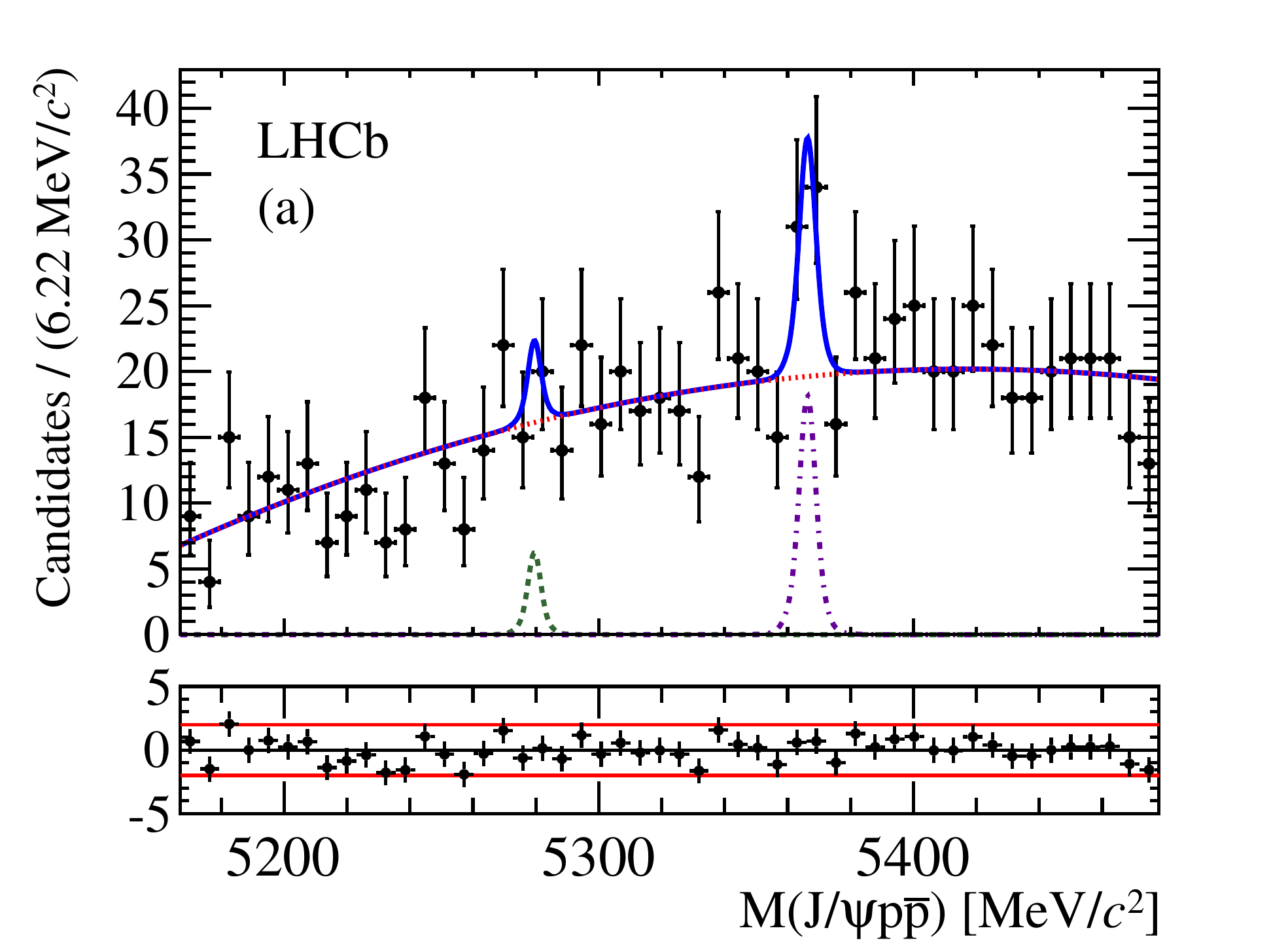}
\includegraphics*[width=0.49\textwidth]{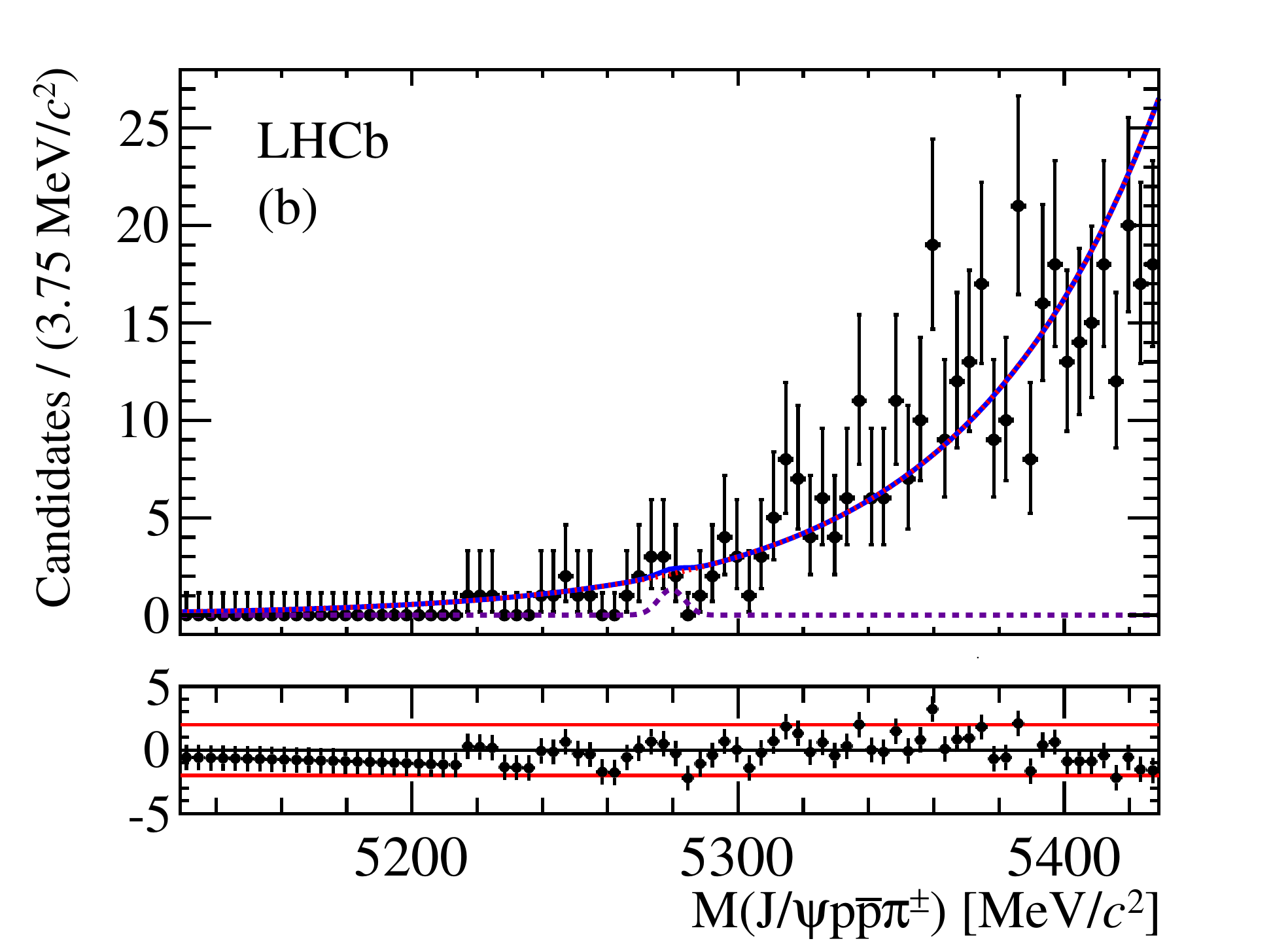}
\end{center}
\caption{\small
Invariant mass distribution of (a) \BToJpsipp and (b) \BuToJpsipppi  
candidates after the full selection.
Each component of the fit model is displayed on the plot:
the signal PDFs are represented by the dot-dashed violet and dashed green line;
the combinatorial background by the dotted red line; 
and the overall fit is given by the solid blue line.
The fit pulls are also shown, with the red lines corresponding to $2\,\sigma$.
The \BuToJpsipppi yield is multiplied by five in order to make the signal position visible. 
}
\label{fig:fit_pph}
\end{figure}

\begin{figure}[htbp]
\begin{center}
\includegraphics*[width=0.49\textwidth]{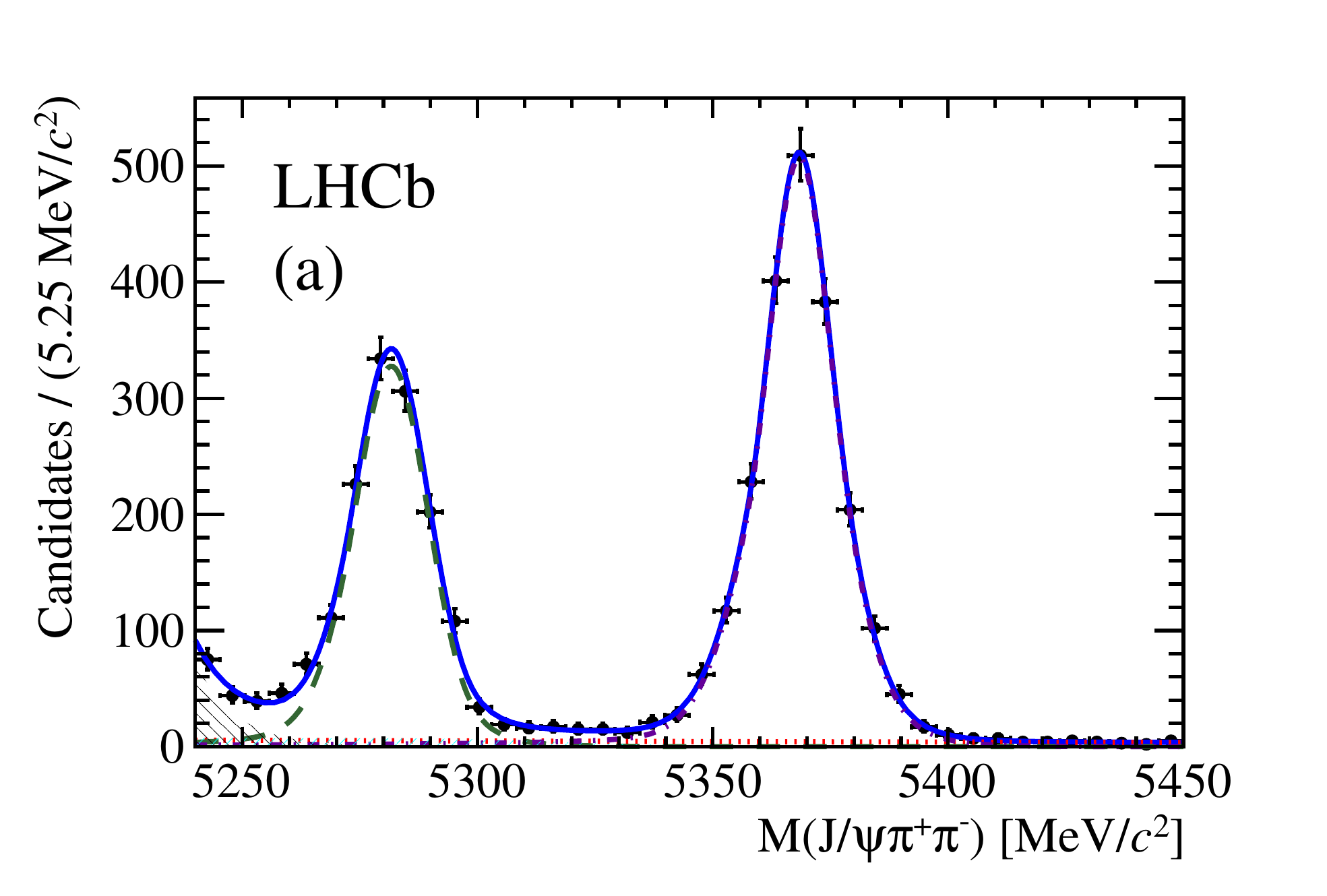}
\includegraphics*[width=0.49\textwidth]{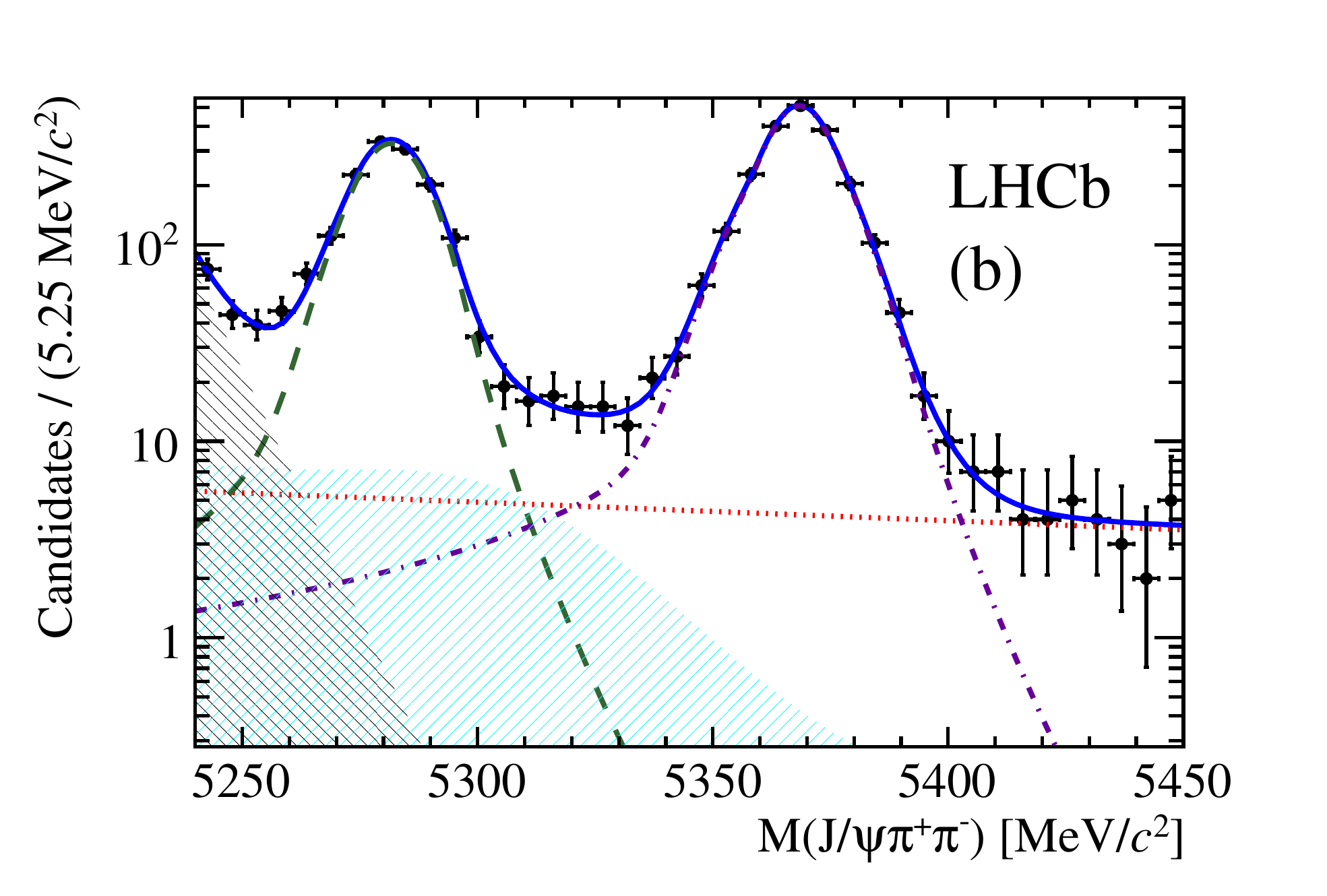}
\includegraphics*[width=0.49\textwidth]{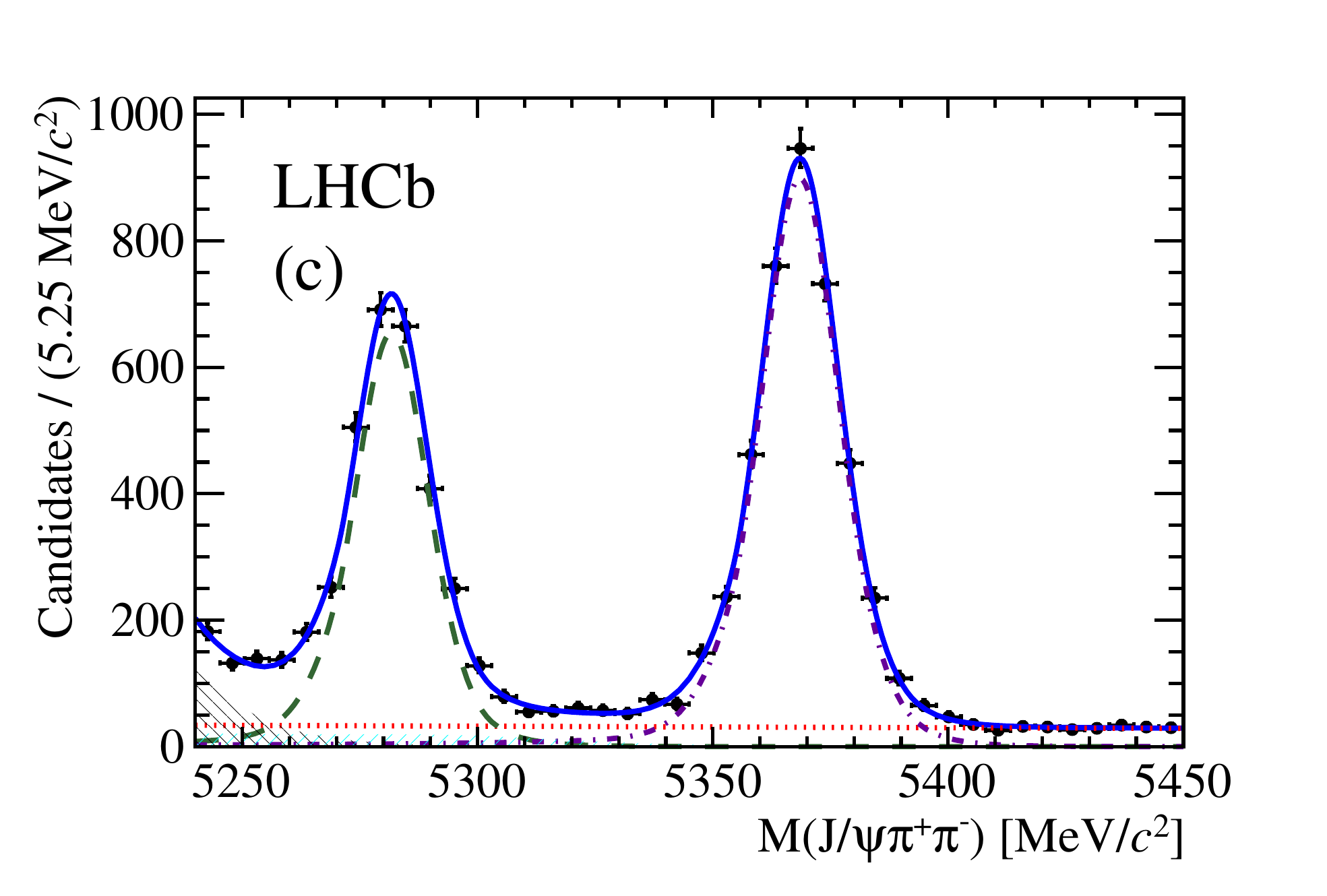}
\includegraphics*[width=0.49\textwidth]{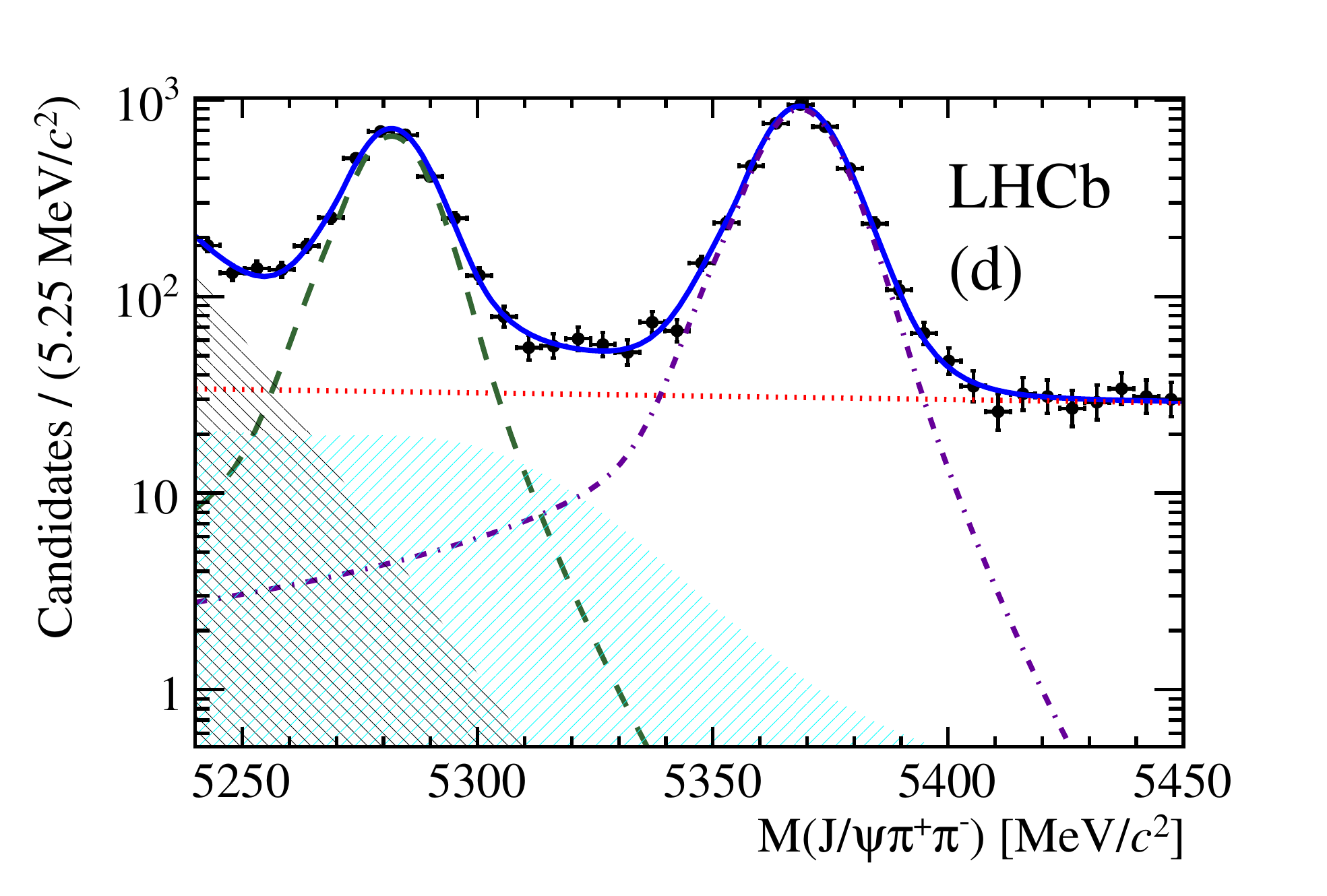}
\end{center}
\caption{\small
Invariant mass distribution of \BToJpsipipi candidates after the full
selection for the (a) \BToJpsipp and
(c) \BuToJpsipppi searches.
The corresponding logarithmic plots are shown in (b) and (d). 
Each component of the fit is represented on the plot:
\BdToJpsipipi signal (green dashed), 
\BsToJpsipipi signal (violet dot-dashed),
\BdToJpsiKpi background (black falling hashed),
${\decay{\Bs}{\jpsi \eta'}}$ background (cyan rising hashed),
and combinatorial background (red dotted).
The overall fit is represented by the solid blue line.
}
\label{fig:fit_pipi}
\end{figure}

The fits to the data are shown in Figs.~\ref{fig:fit_pph} and~\ref{fig:fit_pipi}.
The signal yields are 
$N(\BdToJpsipp)   =  5.9\,^{\,+5.9}_{\,-5.1} \pm 2.5$,
$N(\BsToJpsipp)   = 21.3\,^{\,+8.6}_{\,-7.8} \pm 2.6$ and 
$N(\BuToJpsipppi) =  0.7\,^{\,+3.2}_{\,-2.5} \pm 0.7$,
where the first uncertainties are statistical and the second are systematic 
and are described in the next section.
The numbers of events in the \BsToJpsipipi normalisation channel 
are found to be $2120 \pm 50$ and $4021 \pm 76$ (statistical uncertainties only) when applying the  
selection requirements for the \BToJpsipp and \BuToJpsipppi measurements, respectively.

\begin{figure}[htbp]
\begin{center}
\includegraphics*[width=0.49\textwidth]{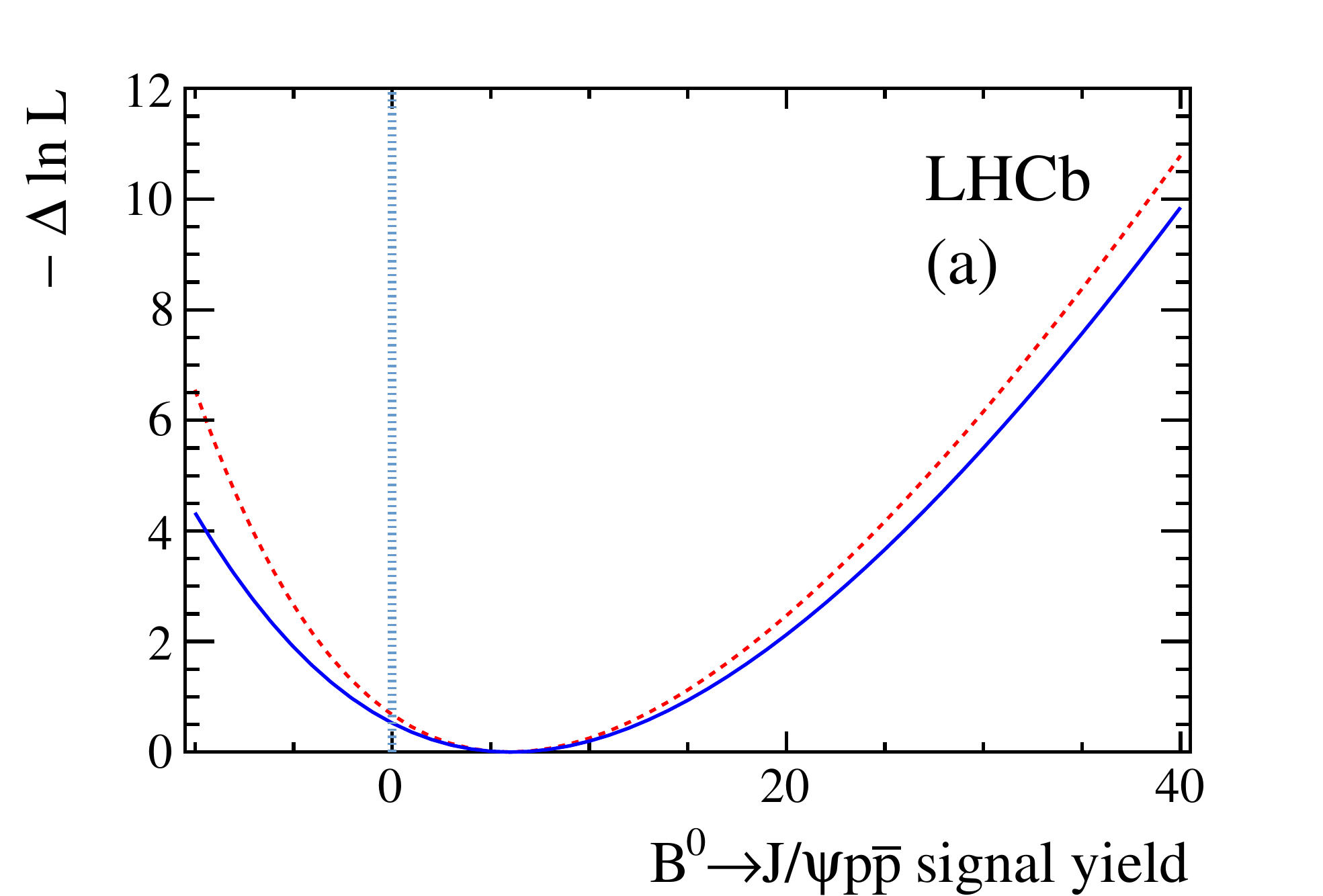}
\includegraphics*[width=0.49\textwidth]{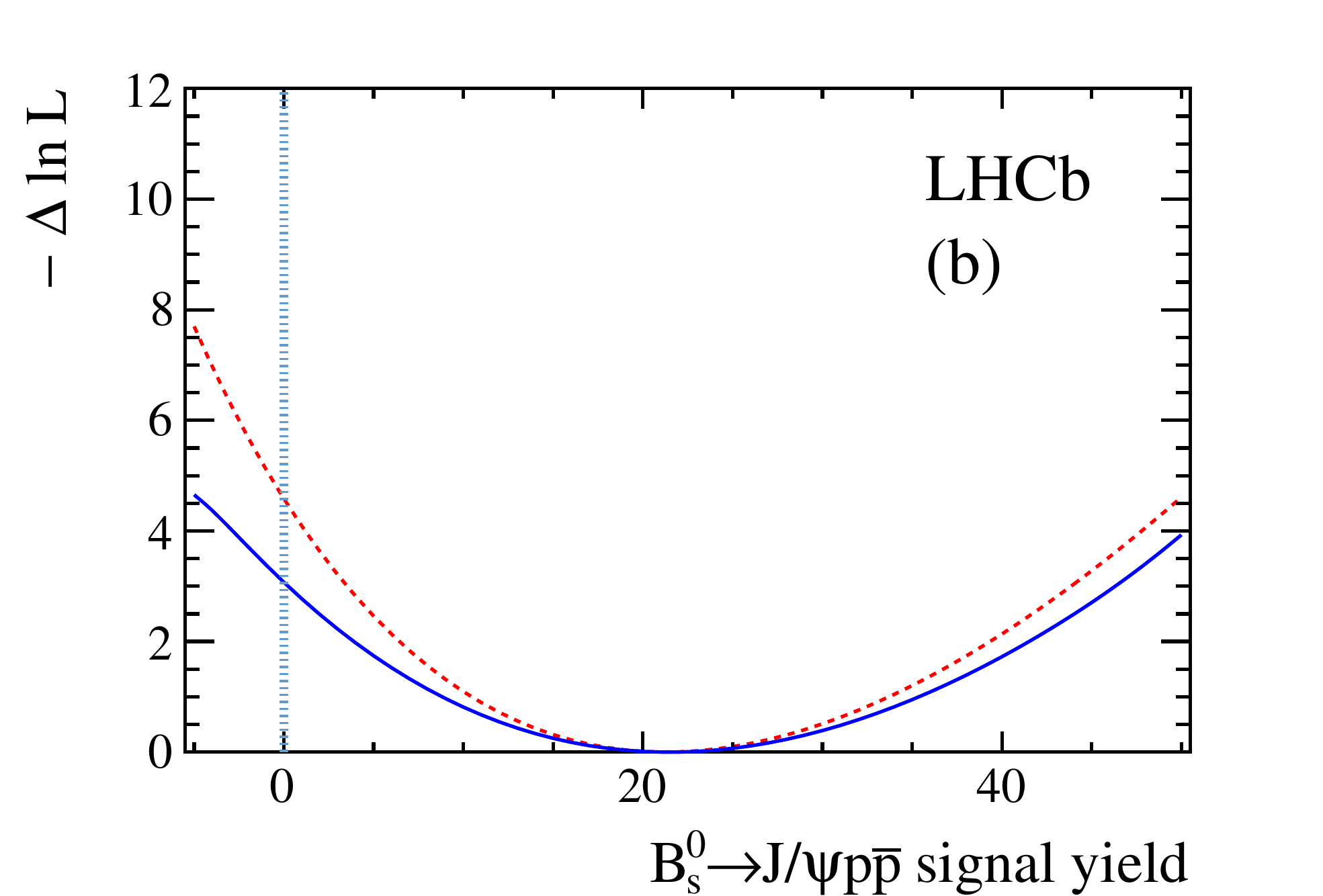}
\includegraphics*[width=0.49\textwidth]{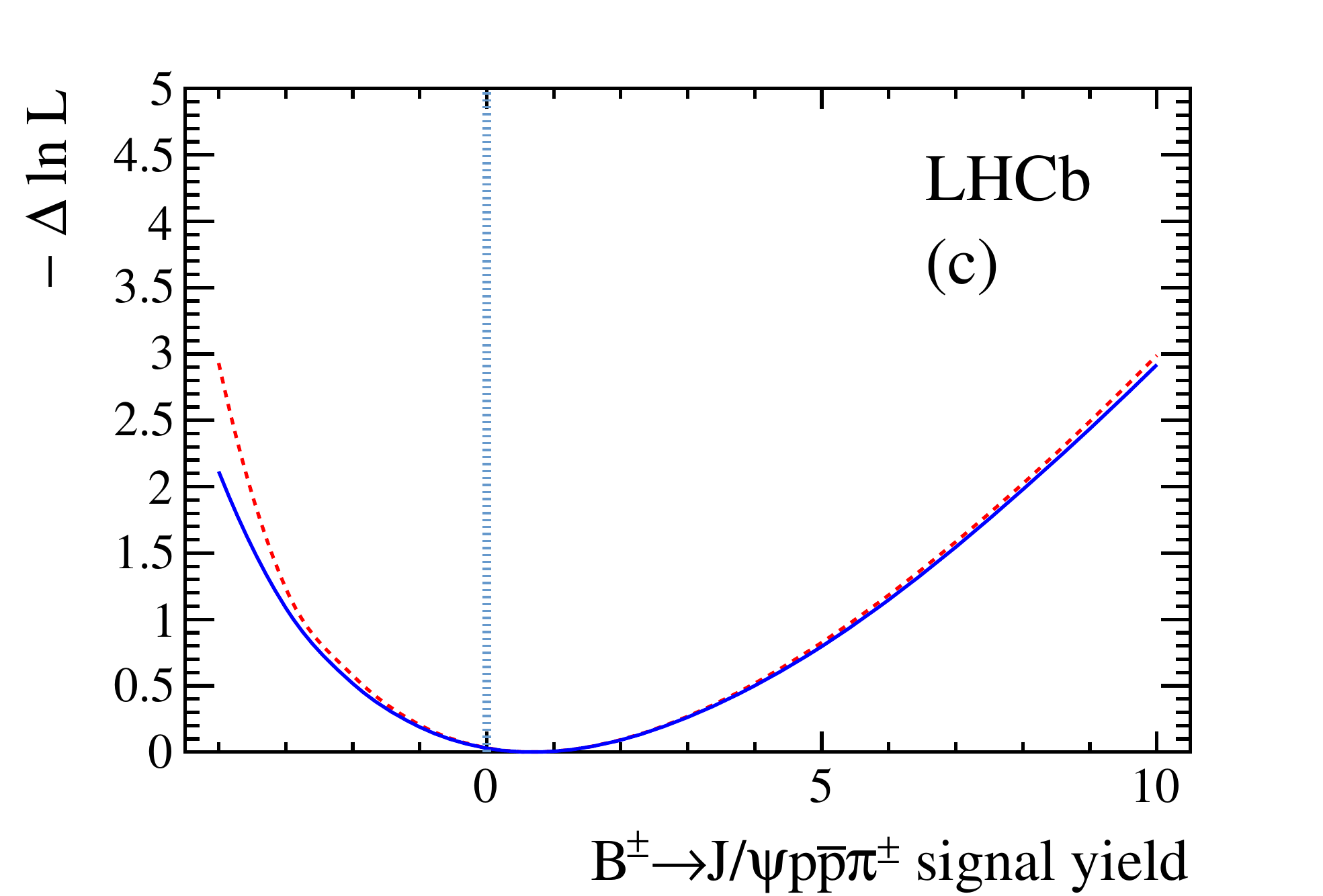}
\end{center}
\caption{\small
Negative log-likelihood profiles for the (a) \BdToJpsipp, (b)
\BsToJpsipp, and (c) \BuToJpsipppi signal yields.
The red dashed line corresponds to the statistical-only profile while the blue line
includes all the systematic uncertainties.
}
\label{fig:profile_likelihood}
\end{figure}

The statistical significances of the signal yields are computed from the
change in the fit likelihood when omitting the corresponding component,
according to $\sqrt{ 2 \ln (L_{\rm sig}/L_{0})}$, where $L_{\rm sig}$ and
$L_{0}$ are the likelihoods from the nominal fit and from the fit omitting
the signal component, respectively.
The statistical significances are found to be $1.2\,\sigma$, $3.0\,\sigma$ and
$0.2\,\sigma$ for the decays \BdToJpsipp, \BsToJpsipp and \BuToJpsipppi, respectively.
The statistical likelihood curve is convolved 
with a Gaussian function of width given by the systematic uncertainty.
The resulting negative log likelihood profiles are shown in Fig.~\ref{fig:profile_likelihood}. 
The total significances of each signal are found to be $1.0\,\sigma$,
$2.8\,\sigma$ and $0.2\,\sigma$ for the modes \BdToJpsipp, \BsToJpsipp and \BuToJpsipppi,
respectively.

\section{Systematic uncertainties}\label{sec:syst}

Many potential sources of systematic uncertainty are reduced by the
choice of the normalisation channel.
Nonetheless, some factors remain that could still affect the measurements of
the branching fractions.
The sources and their values are summarised in Table~\ref{tab:systematics}.

Precise knowledge of the selection efficiencies for the modes is
limited both by the simulation sample size and by the variation of the
efficiency over the multi-body phase space, combined with the
unknown distribution of the signal over the phase space.
The simulation sample size contributes an uncertainty of approximately $1\,\%$ 
in each of the channels, and the effect of efficiency variation across 
the phase space, determined from the spread of values obtained in bins of the  
relevant variables, is evaluated to be $17\,\%$, $14\,\%$ and $23\,\%$ 
for \BdToJpsipp, \BsToJpsipp and \BuToJpsipppi decays, respectively.
The large systematic uncertainties reflect the unknown distribution of signal events across the phase space.
In contrast, the uncertainty for the \BsToJpsipipi normalisation channel is
estimated by varying the binning scheme in the phase space variables and is
found to be only $1\%$ for both the \BToJpsipp and \BuToJpsipppi MVA selections.
Possible biases due to training the MVA using the control channel were investigated
and found to be negligible. 

The proton PID efficiency is measured using a high-purity data sample of ${\decay{\L}{\proton\pim}}$ decays.
By repeating the method with a simulated control sample,
and considering the difference with the simulated signal sample,  
the associated systematic uncertainties are found 
to be $3\,\%$, $3\,\%$ and $2\,\%$ for the modes \BdToJpsipp, \BsToJpsipp and \BuToJpsipppi, respectively.  
Furthermore, the limited sample sizes give an additional $1\,\%$ uncertainty.
In the \BuToJpsipppi channel there is an additional source of uncertainty
due to the different reconstruction efficiencies for the extra pion track
in data and simulation, which is determined to be less than $2\,\%$.

The effect of approximations made in the fit model is investigated 
by considering alternative functional forms for the various signal and
background PDFs. 
The nominal signal shapes are replaced with a bifurcated Gaussian function with asymmetric exponential tails.
The background is modelled with an exponential function for \BToJpsipp decays, whereas a second-order polynomial function 
is used for \BuToJpsipppi and the normalisation channel. 
Combined in quadrature, these sources change the fitted 
yields by $2.5$, $2.6$ and $0.7$ events, which correspond to $42\,\%$,
$12\,\%$  and $92\,\%$ for the \BdToJpsipp, \BsToJpsipp and \BuToJpsipppi 
modes, respectively.
The bias on the determination of the fitted yield is studied with
pseudo-experiments.  
No significant bias is found, and the associated systematic uncertainty is 
$0.2$, $0.3$ and $0.2$ events ($4\,\%$, $1\,\%$ and $26\,\%$) 
for the \BdToJpsipp, \BsToJpsipp and \BuToJpsipppi modes, respectively.

Since a \Bs meson decay is used for the normalisation, the results for 
${\cal B}(\BdToJpsipp)$ and ${\cal B}(\BuToJpsipppi)$ rely on the knowledge 
of the ratio of the fragmentation fractions, 
measured to be $f_s/f_d = 0.256 \pm 0.020$~\cite{LHCb-PAPER-2012-037},
introducing a relative uncertainty of $8\,\%$.
It is assumed that $f_{\uquark} = f_{\dquark}$.
The uncertainty on the measurement of the \BsToJpsipipi 
branching fraction includes a contribution from this source. 
Hence, to avoid double counting, it is omitted when evaluating the systematic uncertainties on the absolute branching fractions.

A series of cross-checks are performed to test the stability of the fit
result.
The PID and MVA requirements are tightened and loosened.
The fit range is restricted to $[5229, 5416]\mevcc$ and  $[5129,
5379]\mevcc$ for \BToJpsipp and \BuToJpsipppi decays, respectively.
No significant change in the results is observed in any of the cross-checks.

\begin{table}[!htb]
 \caption{\small
   Systematic uncertainties on the branching fraction ratios of the decays 
   \BdToJpsipp, \BsToJpsipp and \BuToJpsipppi measured relative to \BsToJpsipipi.
   The total is obtained from the sum in quadrature of all contributions. }
 \label{tab:systematics}
 \centering
 \begin{tabular}{lccc}
   \hline
   Source                  & \multicolumn{3}{c}{Uncertainty on the branching fraction ratio (\%)} \\ 
                               & \BdToJpsipp         & \BsToJpsipp              & \BuToJpsipppi              \\
   \hline
   Event selection             & $\phantom{1}{1}$    & $\phantom{1}{1}$         & $\phantom{1}{1}$           \\
   Efficiency variation        & $17$                & $14$                     & $23$                       \\
   PID simulation sample size  & $\phantom{1}{1}$    & $\phantom{1}{1}$         & $\phantom{1}{1}$           \\
   PID calibration method      & $\phantom{1}{3}$    & $\phantom{1}{3}$         & $\phantom{1}{2}$           \\
   Tracking efficiency         & ---                 & ---                      & $\phantom{1}{2}$           \\
   Fit model                   & $42$                & $12$                     & $92$  \\
   Fit bias                    & $\phantom{1}{4}$    & $\phantom{1}{1}$         & $26$  \\
   Fragmentation fractions     & $\phantom{1}{8}$    & ---                      & $\phantom{1}{8}$           \\
   \hline
   Total                       & $46$                & $19$                     & $98$                       \\
   \hline
 \end{tabular}
\end{table}

\section{Results and conclusions}

The relative branching fractions are determined according to
\begin{equation}\label{eq:master}
\frac{{\cal{B}}({\decay{\B_{q}}{\jpsi \ppbar (\pip)}})}{{\cal{B}}(\BsToJpsipipi)} =   
\frac{\epsilon^{\rm sel}_{\BsToJpsipipi}}{\epsilon^{\rm sel}_{{\decay{\B_{q}}{\jpsi \ppbar (\pip)}}}}
\times \phantom{=}  \frac{\epsilon^{\rm PID}_{\BsToJpsipipi}}{\epsilon^{\rm PID}_{{\decay{\B_{q}}{\jpsi \ppbar (\pip)}}}}
\times \frac{N_{{\decay{\B_{q}}{\jpsi \ppbar (\pip)}}} }{ N_{\BsToJpsipipi}}  
\times \frac{f_{\squark}}{f_{{\quark}}}, 
\end{equation} 
where $\epsilon^{\rm sel}$ is the selection efficiency,
$\epsilon^{\rm PID}$ is the particle identification efficiency,
and $N$ is the signal yield.
The results obtained are
\begin{alignat*}{5}\label{eq:rel_BR_Bd}
\frac{{\cal{B}}(\BdToJpsipp)}{{\cal{B}}(\BsToJpsipipi)}   &\,=\hspace{0.2cm}&  ( 1.0 &\,^{+1.0} _{-0.9}\, &\pm\,& 0.5)\times 10^{-3}\,, \nonumber \\ 
\frac{{\cal{B}}(\BsToJpsipp)}{{\cal{B}}(\BsToJpsipipi)}   &\,=\hspace{0.2cm}&  ( 1.5 &\,^{+0.6} _{-0.5}\, &\pm\,& 0.3)\times 10^{-2}\,, \nonumber  \\ 
\frac{{\cal{B}}(\BuToJpsipppi)}{{\cal{B}}(\BsToJpsipipi)} &\,=\hspace{0.2cm}&  ( 0.27 &\,^{+1.23} _{-0.95}\, &\pm\,& 0.26)\times 10^{-3}\,, \nonumber
\end{alignat*} 
where the first uncertainty is statistical and the second is systematic.
The absolute branching fractions are calculated using the measured branching
fraction of the normalisation channel
${\cal B}(\BsToJpsipipi) = (1.98 \pm 0.20) \times 10^{-4}$~\cite{LHCb-PAPER-2012-005}
\begin{alignat*}{7}
{\cal{B}} (\BdToJpsipp)   &\,=\hspace{0.2cm}&  ( 2.0 &\,^{+1.9} _{-1.7} \stat &\pm\,& 0.9\syst &\pm\,& 0.1\,[\rm{norm}] ) \times 10^{-7}, \nonumber \\
{\cal{B}} (\BsToJpsipp)   &\,=\hspace{0.2cm}&  ( 3.0 &\,^{+1.2} _{-1.1} \stat &\pm\,& 0.6\syst &\pm\,& 0.3\,[\rm{norm}] ) \times 10^{-6}, \nonumber \\ 
{\cal{B}} (\BuToJpsipppi) &\,=\hspace{0.2cm}&  ( 0.54 &\,^{+2.43} _{-1.89} \stat &\pm\,& 0.52\syst &\pm\,& 0.03\,[\rm{norm}] ) \times 10^{-7}, \nonumber
\end{alignat*}
where the third uncertainty originates from the control channel branching fraction measurement. 
The dominant uncertainties are statistical, while the most significant
systematic come from the fit model and from the variation of the efficiency
over the phase space.

Since the significances of the signals are below $3\,\sigma$, 
upper limits at both $90\,\%$ and $95\,\%$ confidence levels (CL) are 
determined using a Bayesian approach, 
with a prior that is uniform in the region with positive branching fraction.
Integrating the likelihood (including all systematic uncertainties), the upper limits are found to be
\begin{eqnarray}
\frac{{\cal{B}} (\BdToJpsipp)}{{\cal{B}}(\BsToJpsipipi)}    &  <  & 2.6~(3.0) \times 10^{-3}~~~{\rm at}~~90\,\%\ (95\,\%)~{\rm CL}\,, \nonumber \\
\frac{{\cal{B}} (\BsToJpsipp)}{{\cal{B}}(\BsToJpsipipi)}    &  <  & 2.4~(2.7) \times 10^{-2}~~~{\rm at}~~90\,\%\ (95\,\%)~{\rm CL}\,, \nonumber \\ 
\frac{{\cal{B}} (\BuToJpsipppi)}{{\cal{B}}(\BsToJpsipipi)}  &  <  & 2.5~(3.1) \times 10^{-3}~~~{\rm at}~~90\,\%\ (95\,\%)~{\rm CL}\,, \nonumber 
\end{eqnarray}
and the absolute limits are
\begin{eqnarray}
{\cal{B}} (\BdToJpsipp)   &  <  & 5.2~(6.0) \times 10^{-7}~~~{\rm at}~~90\,\%\ (95\,\%)~{\rm CL}\,, \nonumber \\
{\cal{B}} (\BsToJpsipp)   &  <  & 4.8~(5.3) \times 10^{-6}~~~{\rm at}~~90\,\%\ (95\,\%)~{\rm CL}\,, \nonumber \\ 
{\cal{B}} (\BuToJpsipppi) &  <  & 5.0~(6.1) \times 10^{-7}~~~{\rm at}~~90\,\%\ (95\,\%)~{\rm CL}\,. \nonumber 
\end{eqnarray}

In summary, using the data sample collected in 2011 by the LHCb experiment
corresponding to an integrated luminosity of $1.0 \invfb$ of $pp$ collisions at $\sqrt{s}=7\tev$, searches for the decay modes \BdToJpsipp, \BsToJpsipp and
\BuToJpsipppi are performed.
No significant signals are seen, and upper limits on the branching
fractions are set.  
A significant improvement in the existing limit for \BdToJpsipp decays is
achieved and first limits on the branching fractions of \BsToJpsipp
and \BuToJpsipppi decays are established.
The limit on the \BdToJpsipp branching fraction is in tension
with the theoretical prediction~\cite{Chen:2008sw}.
The significance of the \BsToJpsipp signal is $2.8\,\sigma$, which motivates
new theoretical calculations of this process as well as improved
experimental searches using larger datasets.

\section*{Acknowledgements}

\noindent We express our gratitude to our colleagues in the CERN
accelerator departments for the excellent performance of the LHC. We
thank the technical and administrative staff at the LHCb
institutes. We acknowledge support from CERN and from the national
agencies: CAPES, CNPq, FAPERJ and FINEP (Brazil); NSFC (China);
CNRS/IN2P3 and Region Auvergne (France); BMBF, DFG, HGF and MPG
(Germany); SFI (Ireland); INFN (Italy); FOM and NWO (The Netherlands);
SCSR (Poland); ANCS/IFA (Romania); MinES, Rosatom, RFBR and NRC
``Kurchatov Institute'' (Russia); MinECo, XuntaGal and GENCAT (Spain);
SNSF and SER (Switzerland); NAS Ukraine (Ukraine); STFC (United
Kingdom); NSF (USA). We also acknowledge the support received from the
ERC under FP7. The Tier1 computing centres are supported by IN2P3
(France), KIT and BMBF (Germany), INFN (Italy), NWO and SURF (The
Netherlands), PIC (Spain), GridPP (United Kingdom). We are thankful
for the computing resources put at our disposal by Yandex LLC
(Russia), as well as to the communities behind the multiple open
source software packages that we depend on.

\addcontentsline{toc}{section}{References}
\setboolean{inbibliography}{true}
\bibliographystyle{LHCb}
\bibliography{main,LHCb-PAPER,LHCb-CONF,LHCb-DP}

\end{document}